\documentclass[12pt]{article}

\usepackage{sbc-template}
\usepackage{graphicx,url}
\usepackage[utf8]{inputenc}
\usepackage[nolist]{acronym}
\usepackage[caption=false]{subfig}
\usepackage{multirow}
\usepackage{xcolor}
\usepackage[hidelinks]{hyperref}

\title{Towards a High Efficiency of Native NDN over\\Wi-Fi 6 for the Internet of Vehicles}

\author{Ygor Amaral B. L. de Sena\inst{1,2}, Kelvin Lopes Dias\inst{1}}

\address{Centro de Informática (CIn)\\Universidade Federal de Pernambuco (UFPE)\\Recife -- PE -- Brazil
  \nextinstitute	
  Unidade Acadêmica de Serra Talhada (UAST)\\Universidade Federal Rural de Pernambuco (UFRPE)\\Serra Talhada -- PE -- Brazil
  \email{ygor.amaral@ufrpe.br, kld@cin.ufpe.br}
}

\begin{document} 

\begin{acronym}[ACRONYM] 
% Change the word ACRONYM above to change the acronym column width.
% The column width is equals to the width of the word that you put.
% Read the manual about acronym package for more examples:
%   http://linorg.usp.br/CTAN/macros/latex/contrib/acronym/acronym.pdf

\acroplural{AP}[APs]{Access Points}
\acro{AP}[AP]{Access Point}
\acro{B5G}[B5G]{Beyond 5G}
\acro{CS}[CS]{Content Store}
\acro{FIB}[FIB]{Forward Information Base}
\acro{GI}[GI]{Guard Interval}
\acro{GPS}[GPS]{Global Positioning System}
\acro{IoT}[IoT]{Internet of Things}
\acro{IoV}[IoV]{Internet of Vehicles}
\acro{KPIs}[KPIs]{Key Performance Indicators}
\acro{MAC}[MAC]{Medium Access Control}
\acro{MCS}[MCS]{Modulation and Coding Set}
\acro{NDN}[NDN]{Named Data Networking}
\acro{NFD}[NFD]{NDN Forwarding Daemon}
\acro{PIT}[PIT]{Pending Interest Table}
\acro{SUMO}[SUMO]{Simulation of Urban Mobility}
\acro{VNDN}[VNDN]{Vehicular NDN}
\acro{WLAN}[WLAN]{Wireless Local Area Network}
\acro{WPA}[WPA]{Wi-Fi Protected Acces}
%\acroplural{CHT}[CHTs]{Constraint-Handling Techniques}
\end{acronym}

\maketitle

\begin{abstract}
Named Data Networking (NDN) is a top-notched architecture to deal with content distribution over the Internet. With the explosion of video streaming transmission and future massive Internet of Things and Vehicles (IoT/IoV) traffic, evolving Wi-Fi networks will play an essential role in such ecosystems. However, Native NDN deployment over wireless networks may not perform well. Wi-Fi broadcasts/multicasts result in reduced throughput due to the usage of basic service mode.  Despite recent initial works addressing that issue, further studies and proposals are required to boost the adoption of Native NDN. We advocate that an initial step towards designing a feasible Native NDN over wireless networks should be understanding the challenges in emerging scenarios and providing a uniform baseline to compare and advance proposals. To this end, first, we highlight some challenges and directions to improve throughput and energy efficiency, reduce processing overhead, and security issues. Next, we propose a variant of NDN that minimizes the problems identified by performing transmission via unicast to avoid storms in wireless networks. Finally, we conducted a performance evaluation to compare Standard Native NDN with our proposal on Wi-Fi 6 vehicular networks. The results show that our proposal outperforms the Standard NDN in the evaluated scenarios, reaching values close to 89\% of satisfied requests, achieving more than 200\% of data received than Standard NDN.
\end{abstract}

\section{Introduction}\label{sec:introduction}

\ac{NDN}~\cite{Zhang2014} has been advocated as a top-notched architecture to deal with the content distribution over the Internet. Instead of using the traditional end-to-end IP-based communication, \ac{NDN} adopts a hop-by-hop approach to distributing and retrieving content on the Internet. Thus, \ac{NDN} does not need network layer addressing but relies on names to request the desired content. This solution has several advantages, especially in mobility contexts~\cite{Sena2022}. When it comes to vehicles as end-users, this architecture has been promoted to overcome the intrinsic dynamic and challenging scenarios of wireless networks and, in particular, is well-suited to the \ac{IoV} through different solutions based on \ac{VNDN}.

\ac{NDN} deployments generally follow two approaches: 1) Overlay \ac{NDN}, i.e., running over IP networks, or 2) Native \ac{NDN}, which replaces the IP protocol to run directly over the link-layer technologies~\cite{Nour2019}. Some solutions have been devised to show the benefits of \ac{NDN} in wireless networks with scenarios ranging from video transmission to \ac{IoT} and \ac{IoV} content distribution. Despite existing works on the synergy between \ac{NDN} and \ac{IoV}~\cite{Grassi2014,Anastasiades2016,Coutinho2018,Duarte2019,Wang2020}, the majority of \ac{NDN} networking experiments still run low bandwidth Wi-Fi.

With the explosion of video streaming transmission and future massive \ac{IoT} and \ac{IoV} traffic~\cite{Rothmuller2020}, evolving Wi-Fi networks will play an essential role in such ecosystems. Recently, Wi-Fi has evolved into a new version, known as WiFi 6 (IEEE 802.11ax), which is efficient to transmit at a high transfer rate~\cite{Khorov2019}, and has been an alternative for free of charge / public access complementary technology to 5G connectivity. Furthermore, Wi-Fi 7 seems to address the requirements of \ac{KPIs} aligned with those related to \ac{B5G} cellular networks. Hence, realistic performance evaluation and insights for the \ac{NDN} landscape should be considered to understand synergies and drawbacks when running \ac{NDN} over high throughput Wi-Fi networks. However, when it comes to Native \ac{NDN} deployment over contention-based wireless networks, such as Wi-Fi, \ac{NDN} may not perform well. Wi-Fi broadcast/multicast transmissions result in reduced throughput due to the usage of basic service~\cite{IEEE802.11-2021}.   

Hence, we propose an improvement in the \ac{NDN} architecture, to achieve high transfer rates in wireless networks. Our proposal aims to learn \ac{MAC} addresses in both directions (upstream/downstream), so, as not to perform unnecessary broadcasts at the link layer and thus avoid the basic service of Wi-Fi networks. Our solution only affects transmissions at the link layer. At the network layer, communication is maintained according to the standard. To deploy our solution we use ndnSIM 2~\cite{Mastorakis2017a}, a key tool for evolving \ac{NDN} and flourishment of improvements of its various facets. 

In addition to our proposal and the Standard \ac{NDN}, we also include two intermediate approaches: the former avoids broadcast only in the upstream direction, while the latter one in the downstream. In this way, we conducted a performance evaluation of these four Native \ac{NDN} deployments in an \ac{IoV} context considering three Wi-Fi 6 hotspots, with vehicles performing handovers along the avenue. The vehicular traffic has been modeled based on real traces. Our simulation results show that the proposal of this work outperforms Standard \ac{NDN}, reaching around 89\% of requests satisfied, against only 38\% in the best case of Standard \ac{NDN} deployment.

The remainder of this paper is organized as follows. We discuss in Section~\ref{sec:challenges} the challenges of running Native \ac{NDN} over wireless networks. In Section~\ref{sec:related_work} we describe related work. Section~\ref{sec:proposal} describes
our proposal for mapping layer-2 addresses to Native \ac{NDN}. We address all the details of the simulation-based experiments in Section~\ref{sec:experimental_setup}. We then discuss the performance evaluation results in Section~\ref{sec:performance_evaluation}. Finally, we conclude our paper in Section~\ref{sec:conclusions}.

\section{Challenges of Native \ac{NDN} over Wireless Networks}\label{sec:challenges}
In this section, we describe the main current challenges in using Native \ac{NDN} over wireless networks.

\subsection{Throughput}
Due to the absence of layer-2 address resolution, Native \ac{NDN} does not know the link addresses, thus generating broadcast storms in wireless networks. Besides that, a broadcast transmission uses only the basic service~\cite{IEEE802.11-2021} provided by most 802.11 variants. Hence, throughput is much lower and retransmissions are disabled, providing less reliability. 

One work evaluated that unicast traffic is promising for \ac{NDN} networks~\cite{Kietzmann2017}, but did not propose any address mapping mechanism. Furthermore, some works have proposed mechanisms to make broadcast transmissions more responsive on the wireless channel. Self-learning forwarding strategy has been proposed~\cite{Shi2017} and improved~\cite{Liang2020} to adaptively transmit via multicast or unicast. However, this strategy does not perform \mbox{layer-2} address mapping, and it is not possible to perform unicast with Native \ac{NDN} without prior configuration of the \mbox{layer-2} addresses. A multicast rate adaptation scheme has been proposed \cite{Wu2018}, which performs passive mapping of layer-2 addresses to deliver data packets (downstream), but still broadcasts when sending interest packets. As such, high throughput is still an open issue in Native \ac{NDN} over wireless networks.

\subsection{Processing Overhead}
Since all packet transmissions are performed via layer-2 broadcast, inevitably devices will process multiple packets on the \ac{NFD}~\cite{Afanasyev2018a} unnecessarily. This issue is even more challenging in \ac{IoT} devices, as they have a small amount of computing resources. Furthermore, this processing overhead is energy inefficient.

\subsection{Energy Efficiency}
Energy efficiency is one critical issue for \ac{IoT} devices as they will typically be battery-powered, and the longer lifetime, the better. In addition to the processing overhead issue, broadcast transmissions in wireless networks are costly in terms of energy consumption because they need to acquire the medium for a long time~\cite{rfc8352}. Thus, the more unnecessary broadcast is avoided, the more efficient is the energy consumption.

\subsection{Security}
Wireless networks introduce a number of security issues since spectrum transmissions are broadcast and vulnerable to unauthorized access. Furthermore, any device within radio range can receive and transmit data. Because of this, several \ac{WPA} techniques have been proposed to provide access control, authentication and privacy in the data exchange through temporary keys during the communication between the devices. Thus, even if the communication is transmitted on the radio, the packets are encrypted individually during the nodes communication. In this case, a third node may even have access to the packet, but it will be encrypted with an unknown key, preventing unauthorized access. However, if sensitive data is sent by a device using Native \ac{NDN} through layer-2 broadcast, these \ac{WPA} security measures are compromised since all devices within the wireless network will have access to the transmitted packets. 

\section{Related Work}\label{sec:related_work}
The Standard \ac{NDN} architecture has no layer-2 address resolution. Thus, the Native \ac{NDN} only transmits through broadcast communication, which requires several precautions to avoid storms in wireless networks, since a broadcast transmission uses only the basic service~\cite{IEEE802.11-2021} provided by most 802.11 variants, where throughput is much lower and retransmissions are disabled, providing less reliability. Conversely, the Overlay \ac{NDN}, when running over the IP, can use the existing layer-2 address resolution mechanism. Thus, Overlay \ac{NDN} has name-based routing, with hop-by-hop communication, but without the need for all transmissions to occur through broadcast~\cite{Sena2022}.

Some works have proposed mechanisms to make broadcast transmissions more responsive. An approach called NLB~\cite{Li2015a} has been proposed for efficient live video broadcasting over Overlay \ac{NDN} in wireless networks. NLB is a leader-based mechanism to suppress duplicate requests, where a single consumer requests (via UDP unicast) and everyone receives the same data (via UDP broadcast). A multicast rate adaptation scheme in wireless networks has been proposed in~\cite{Wu2018}. With this approach, interests are always sent via layer-2 broadcast. However, a mapping mechanism between the \ac{PIT} entry and the layer-2 address has been developed that allows the sending of data via layer-2 unicast. In this way, the proposed scheme can decide when it is better to send data packets via unicast or broadcast.

A broadcast-based adaptive forwarding strategy called self-learning has been proposed~\cite{Shi2017} and improved~\cite{Liang2020} to learn paths without needing routing algorithms. This is useful in wireless networks where nodes can be mobile and routes can change dynamically. To learn routes, the strategy broadcasts the first interest and upon receiving the data, it learns which paths have the content with the respective prefix. This way, the next interests can be sent via unicast to the learned paths. \mbox{DQN-AF}~\cite{Sena2020} is also an adaptive forwarding strategy that, through deep reinforcement learning, forwards through the best paths. However, these approaches do not perform layer-2 address mapping, and it is not possible to perform unicast with Native \ac{NDN}, without prior configuration. The experiment in~\cite{Liang2020} were performed with Overlay \ac{NDN}, using UDP unicast and UDP broadcast.

In the specific context of \ac{VNDN},~\cite{Grassi2014} performed a study in which all packets are sent via layer-2 broadcast. However, to reduce the disadvantages of exhaustive broadcast, the authors created a mechanism that uses \ac{GPS} information to perform forwarding based on distance, avoiding two nearby cars from sending packets simultaneously to use the wireless channel more efficiently. Moreover, to restrict the spread of interest packets, a hop limitation has been applied.

The dynamic unicast~\cite{Anastasiades2016} is a routing protocol devised to perform an implicit content discovery through broadcast transmissions and dynamic content retrieval with efficient unicast links, without the need for location information. When a unicast path is broken, it can be reestablished when new interests are sent via broadcast by neighboring nodes. Another protocol, called LOCOS~\cite{Coutinho2018}, has been proposed for content discovery and retrieval in \ac{VNDN}. LOCOS performs a directed search for content based on the location. Once the producer changes their location, requests cannot be satisfied until the new location is discovered. The protocol will periodically conduct a controlled search in the vicinity area to find the new location through transmissions of interests via broadcast. In this way, LOCOS reduces the storm problem while forwarding is directed to the nearest source.

MobiVNDN~\cite{Duarte2019} is a variant of the \ac{NDN} for \ac{VNDN} and has been proposed to mitigate the performance problems of \ac{VNDN} in wireless networks. In this proposal, the interest and data packets have some differences from the Standard \ac{NDN}. Moreover, a new packet called advertisement has been proposed to propagate content availability. In MobiVNDN, vehicles exchange location and speed information with each other to assist in forwarding and calculating the probability of communication interruptions. In this approach, the geographical location provided by the GPS also performs a key role in preventing unnecessary use of the wireless channel and thus minimizing the problems of broadcast storms. Still, even though MobiVNDN makes better use of the wireless channel, communication is also done through broadcast at layer 2.

An approach has been proposed~\cite{Wang2020} to improve data delivery on \ac{VNDN} with a scheme in which the vehicular backbone has a unicast data delivery process. Despite a small scenario with few vehicular nodes, the simulation results show an increase in efficiency and the authors conclude that unicast is one of the responsible for reducing communication costs in wireless networks.

\section{Improving Native NDN}\label{sec:proposal}
In this section, we detail the solution proposed in this work. The standard architecture of \ac{NDN} does not have any layer-2 address mapping mechanism. It is a premise of the designers that \ac{NDN} should not manipulate addresses. Such an approach is very promising at the network layer, as the routing algorithms will not be based on device location, which is particularly useful in the context of \ac{IoV} and \ac{IoT} mobility. However, we advocate that \ac{NDN} must handle layer-2 addresses, especially in contention-based wireless networks such as Wi-Fi 6. The problem is that these wireless network technologies change broadcast transmissions to the basic service~\cite{IEEE802.11-2021}, as they suffer from broadcast storms, degrading transmission rates and retransmissions are disabled, providing less reliability.

\begin{figure}[!htb]
	\centering
	\includegraphics[width=0.6\textwidth]{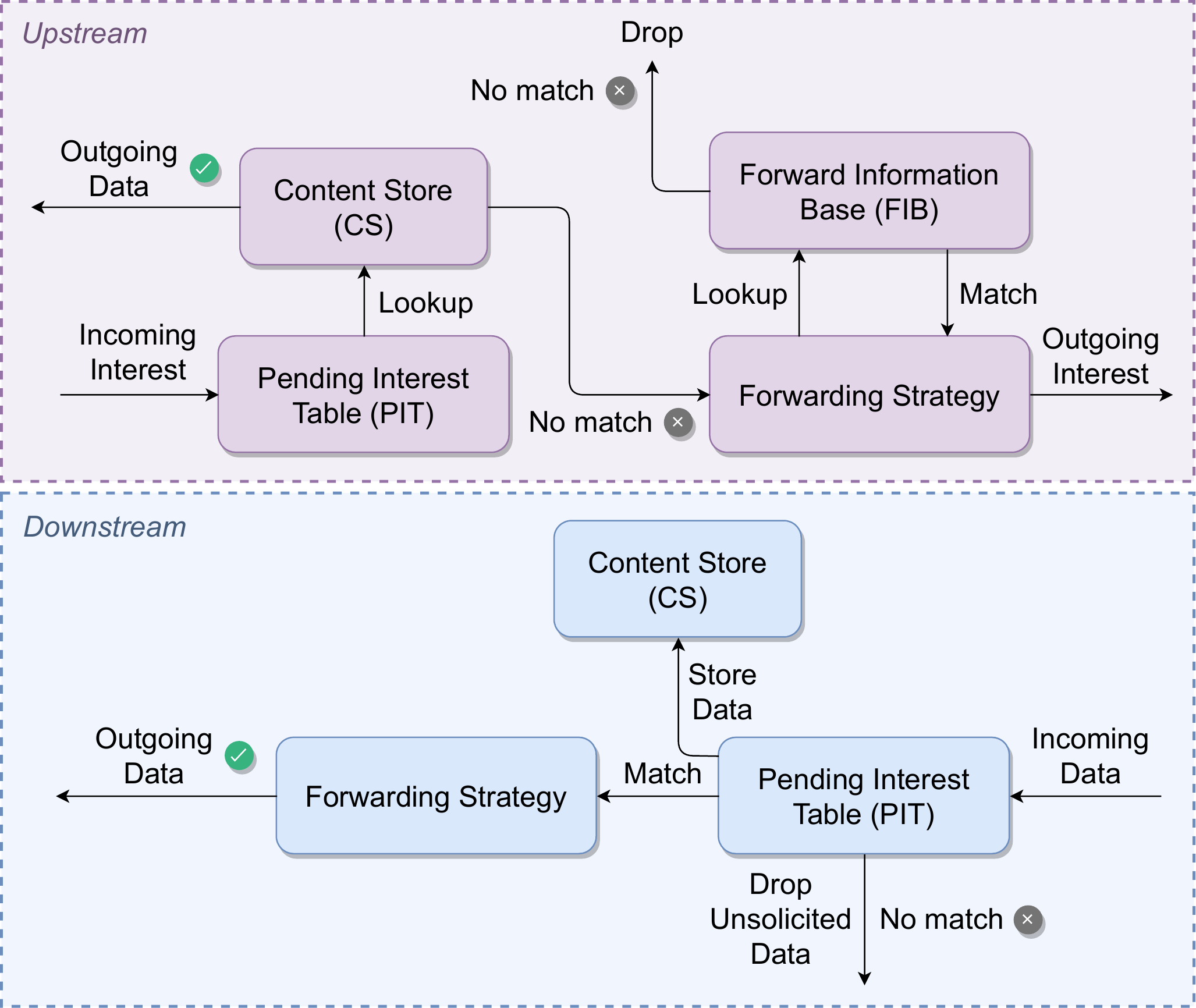}
	\caption{NFD packet forwarding.}
	\label{fig:ndn_forwarding}
\end{figure}

\begin{figure}[!htb]
	\centering
	\includegraphics[width=0.8\textwidth]{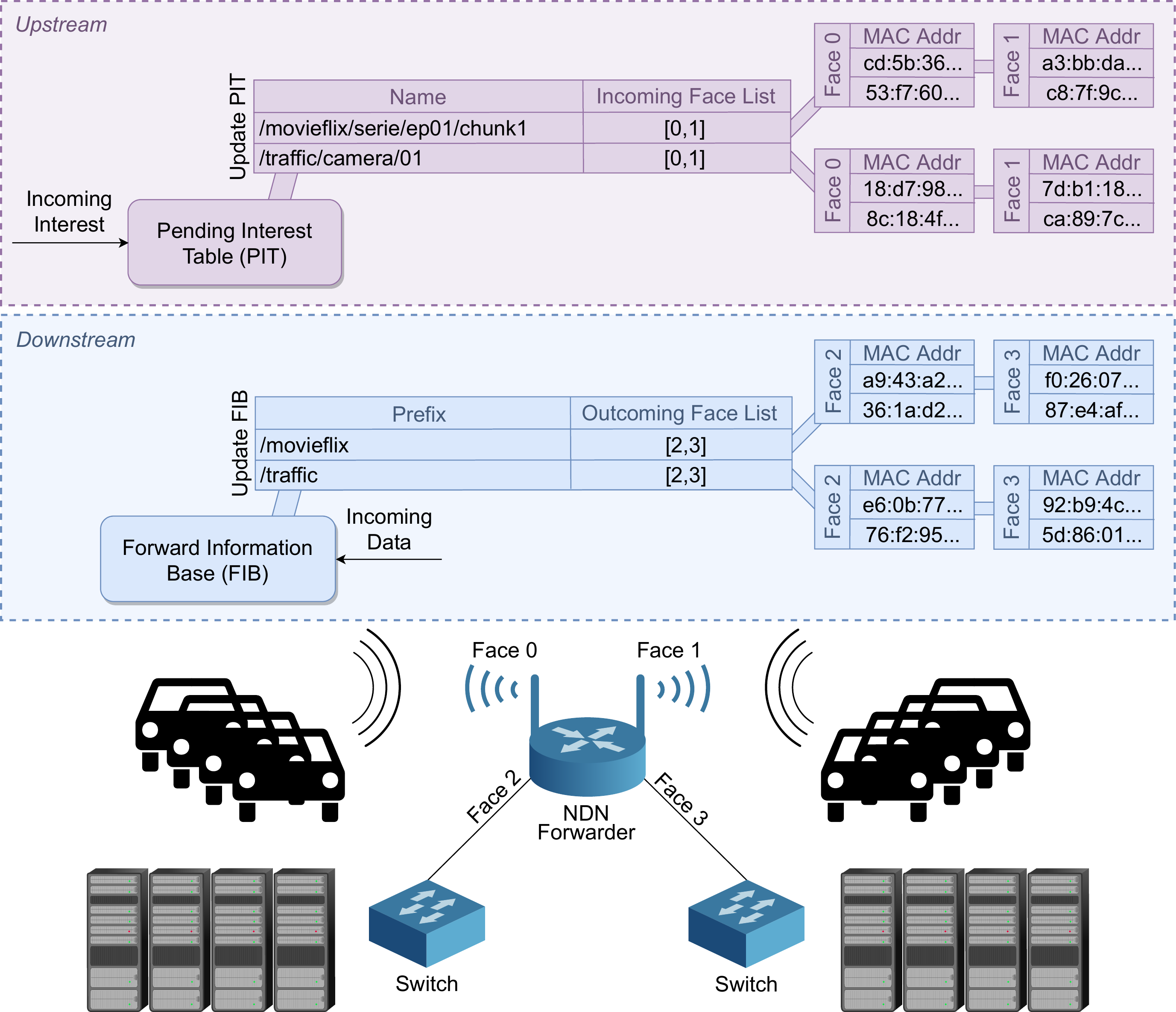}
	\caption{NDN with changes to learn \ac{MAC} addresses.}
	\label{fig:ndn_mac_learning}
\end{figure}

These popular wireless technologies are unlikely to change the way they operate so that \ac{NDN} can exhaustive transmit via broadcast. Therefore, we propose a passive layer-2 address mapping mechanism, without changing any \ac{NDN} message exchange behavior at the network layer. We implemented this mechanism in the official \ac{NDN} forwarding software, known as \ac{NFD}. The basic functioning of the forwarding performed by the \ac{NFD} is shown in Figure~\ref{fig:ndn_forwarding} and our proposed mechanism is shown in Figure~\ref{fig:ndn_mac_learning}.

By default, when an interest packet arrives in the \ac{NFD} (upstream), it is initially registered in the \ac{PIT} and if it does not find the desired data in the \ac{CS}, the interest packet will be forwarded to the next hop through the Forwarding Strategy, querying routing information in the \ac{FIB}. In our proposal, we included the \ac{MAC} address of the sender of interest in the Incoming Face List, as shown in Figure~\ref{fig:ndn_mac_learning}. It is important to remember that in a network port, one or more devices can be reachable through that same port, mainly in wireless networks, so that each face may have one or more \ac{MAC} addresses associated with it. These addresses saved in the \ac{PIT} will only actually be used when sending the data packet to the interested host.

In the downstream direction, by default, when a data packet arrives at the \ac{NFD}, it is initially checked in the \ac{PIT} if there is at least one device with interest registered for that data. If so, it will be stored in the \ac{CS} and forwarded to the host interested through the Forwarding Strategy. In our proposal, we also included the \ac{MAC} address of the neighbor hop that forwarded the received data packet in the Outcoming Face List, inside the \ac{FIB}, as shown in Figure~\ref{fig:ndn_mac_learning}. Similar to the addresses saved in the \ac{PIT}, each face mapped to the \ac{FIB} may have one or more \ac{MAC} addresses associated with it as well. These addresses saved in the \ac{FIB} will only be used when new packets of interest are forwarded to the next hop.

With this proposed mechanism, a \ac{NDN} node passively learns the \ac{MAC} addresses of its neighbors, being able to perform unicast transmissions whenever possible and desirable, to avoid broadcast storms.

\section{Experimental Setup}\label{sec:experimental_setup}

We performed our experiments with ndnSIM 2.8~\cite{Mastorakis2017a}, adding the \ac{NDN} stack to a modified ns-3~\cite{ns3}. However, the version used is still 3.30.1, so we migrated to ns-3.33 due to the new features of the 802.11ax module. Our experiments used four variants of \ac{NDN} deployments (see Table~\ref{tab:list_evaluated_scenarios}) over Wi-Fi 6 networks in the vehicular context.

\begin{table}[htbp]
	\centering
	\caption{List of evaluated Native \ac{NDN} deployment instances}
	\begin{tabular}{cccc}
		\hline Native \ac{NDN} Deployment & Link layer operating mode & Scenarios & Instance\\ 
		\hline
		\hline \multirow{2}{*}{Standard} & \multirow{2}{*}{Broadcast in both directions} & 1 & Standard-1  \\
		\cline{3-4} & & 2 & Standard-2 \\
		\hline \multirow{2}{*}{Up} & \multirow{2}{*}{Unicast upstream only} & 1 & Up-1   \\
		\cline{3-4} & & 2 & Up-2   \\
		\hline \multirow{2}{*}{Down} & \multirow{2}{*}{Unicast downstream only} & 1 & Down-1   \\
		\cline{3-4} & & 2 & Down-2   \\
		\hline \multirow{2}{*}{Proposal} & \multirow{2}{*}{Unicast in both directions} & 1 & Proposal-1   \\
		\cline{3-4} & & 2 & Proposal-2   \\
		\hline
	\end{tabular}
	\label{tab:list_evaluated_scenarios}
\end{table}

\subsection{Vehicular Traffic Modeling}\label{sec:traffic_modeling}
To model vehicular traffic realistically in \ac{SUMO} \cite{Lopez2018}, we collected open data~\cite{cttu2019recife} from the transportation authority of Recife, Brazil, and we chose the data of 2019, as this year vehicular traffic was not influenced by the Covid-19 pandemic. The transport authority provides data such as the date/time and speed of each car traveling the streets for all city traffic sensors. The traffic sensor identified by FS037REC was chosen to have its data analyzed.

We calculate the average traffic on business days and based on this, we model the scenario with 172 meters of avenue, 3 bus stops and 125 vehicles over 300 seconds, with an average and a maximum speed of 31 km/h and 60 km/h, respectively.

\subsection{Scenarios}\label{sec:scenario}
Our simulation scenarios consist of 125 vehicular nodes that, along the 172 meters of the avenue, will be connected through \ac{NDN} \acp{AP} in a Wi-Fi 6 network, 802.11ax standard with \ac{MCS}~11 and 800ns of \ac{GI}.

\begin{figure}[htbp]
	\centerline{\includegraphics[width=0.67\columnwidth]{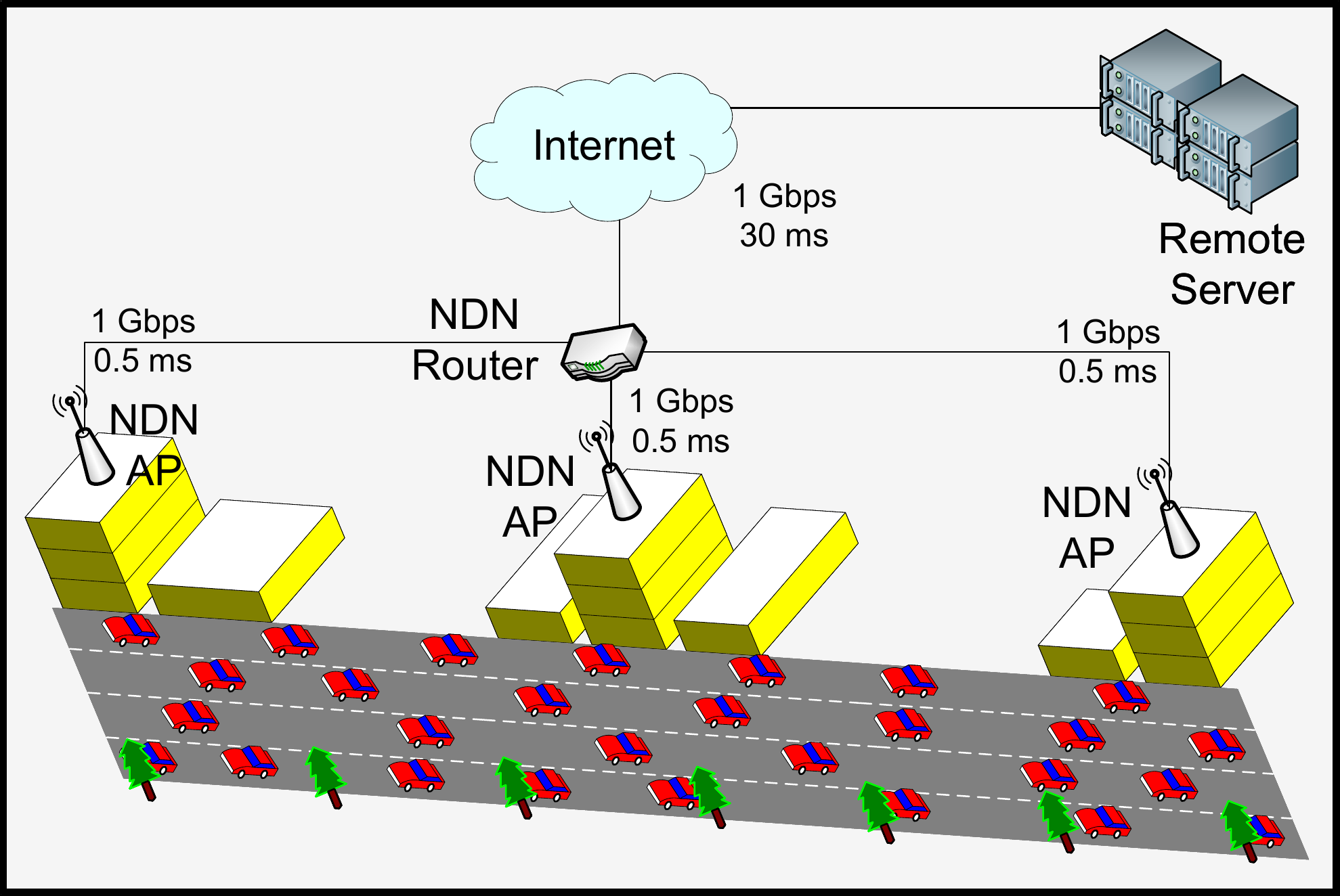}}
	\caption{Vehicular NDN network topology.}
	\label{fig:topology}
\end{figure}

As shown in Figure~\ref{fig:topology}, the \ac{NDN} \acp{AP} are distributed along the avenue. The \ac{NDN} \acp{AP} are connected to an \ac{NDN} router through point-to-point links with 1~Gbps bandwidth with 0.5~ms delay. The \ac{NDN} router it is connected to the remote server (the producer) with a 1~Gbps point-to-point link and 30~ms delay. The \ac{CS} size of the \acp{AP} and router is 10,000 packets and for all other nodes it is 0. The payload of the data packets is 1024 bytes.

In our scenarios, the vehicular nodes send interest packets at a constant rate, defined uniformly between 50 and 100 packets per second for each vehicle. We created two scenarios identified by a suffix 1 or 2 at the end of the name for each instance of deployment (please, see Table~\ref{tab:list_evaluated_scenarios}). In the first, all vehicular nodes use the \textit{ConsumerCbr} application, available by default in the ndnSIM 2. We define that each vehicle requests content with a different prefix, so we force all vehicles to request distinct data between them. Because of this peculiarity, there should be no advantage to broadcast traffic. In the second scenario, we randomly choose 50\% of the vehicles to use the \textit{ConsumerCbr} application in the same way as in the first scenario, and the rest to use the \textit{ModifiedConsumerCbr}, a new application modified by us that sends interest packets with the sequence number based on the simulation time, therefore, vehicles request contents with the same name at the same time. Consequently, many vehicles request the same content and there may be an advantage in broadcast traffic. In these proposed scenarios, we performed 31 simulations for each instance present in Table~\ref{tab:list_evaluated_scenarios}.

\section{Performance Evaluation}\label{sec:performance_evaluation}
In this section, we inform the statistical methods used, as well as present and discuss the results obtained with the simulations performed.

\subsection{Statistical Tests}\label{sec:statistical_tests}

Arcuri and Briand~\cite{Arcuri2011} discuss the usage of statistical testing to analyze randomized algorithms in software engineering. Based on that, we chose Shapiro-Wilk to test the normality of the results. Although the data follow a normal distribution, homoscedasticity is not satisfied, that is, the variances between the distributions are not equivalent. Thus, we chose to use the following statistical tests: Mann-Whitney U-test, a non-parametric significance test; Vargha and Delaney's $\hat{A}_{12}$, a non-parametric effect size test, for assessing whether there are statistical differences among the obtained results. We used a confidence level of 95\% in all cases. All statistical analyses and tests were run using SciPy~\cite{2020SciPy-NMeth}, an open-source scientific tool.

\subsection{Results}\label{sec:results}
This section presents the simulation results of the Native \ac{NDN} deployments presented in Table~\ref{tab:list_evaluated_scenarios}. Our objective is to verify if there is a statistical difference in performance and which approach is the best in our scenarios. We started the discussion with the Mann-Whitney U-test, that deals with their stochastic ranking~\cite{Arcuri2011} to observe the probability that one population will have its values higher than the other and thus verify the statistical significance between these populations. Our null hypothesis ($H_{0}$) is rejected in favor of the alternative hypothesis ($H_{a}$) when the p-value is less than or equal to 0.05, suggesting that the evaluated instances achieved statistically different performances. Otherwise, it suggests that the evaluated instances achieved the same performance.

\begin{figure}[!htb]
	\centering
	\includegraphics[width=0.68\textwidth]{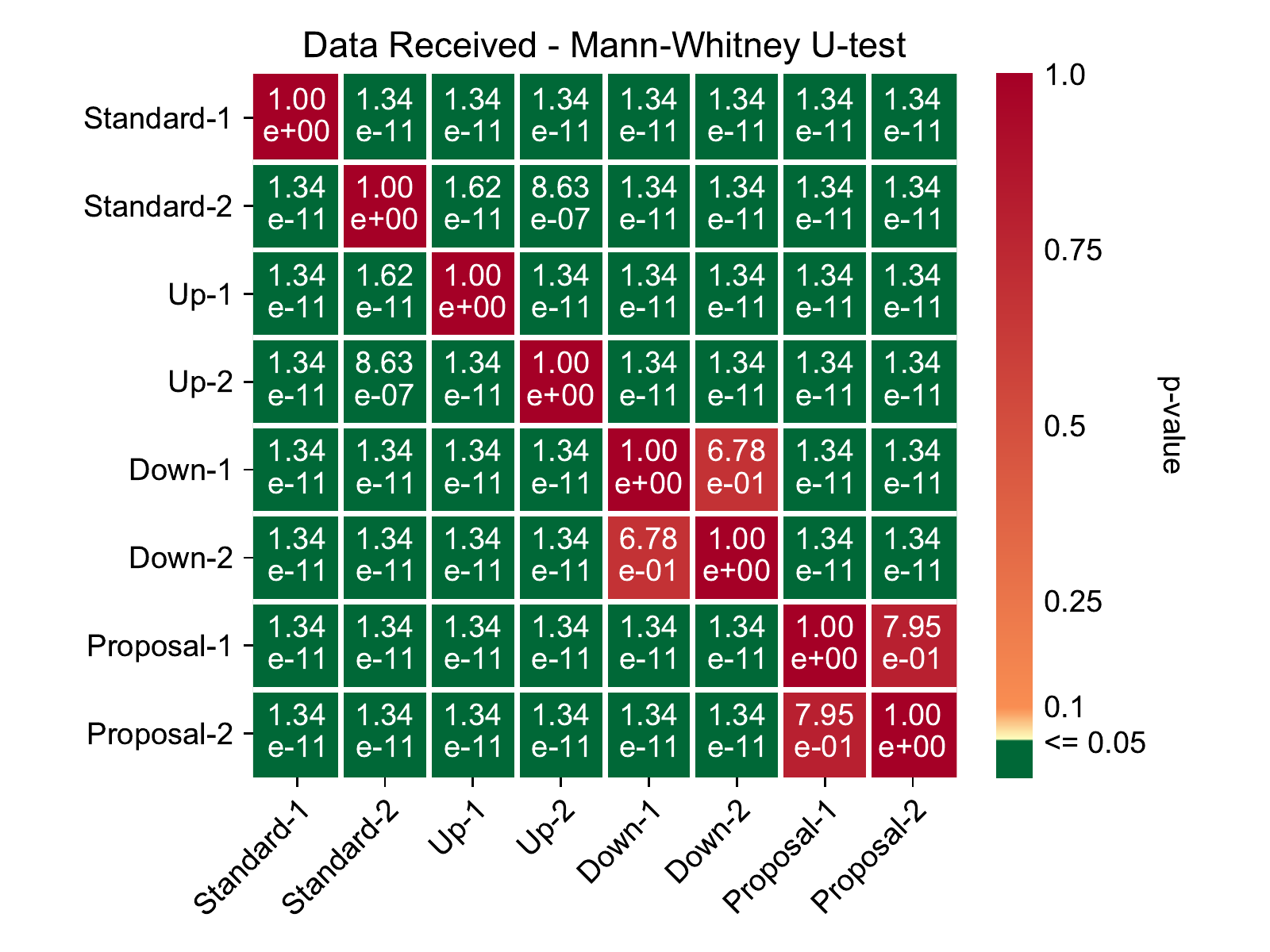}
	\caption{The Mann-Whitney U-test.}
	\label{fig:mw}
\end{figure}

Figure~\ref{fig:mw} shows U-test p-values for the metric of data received in vehicles in each instance evaluated of Table~\ref{tab:list_evaluated_scenarios}. All evaluated Native \ac{NDN} deployments achieved statistically different performances from one another, which confirms that changing the transmission mode from broadcast to unicast changes performance in a NDN deployment over wireless networks. The only times the alternative hypothesis has been rejected was when comparing instances Down-1 with Down-2, and Proposal-1 with Proposal-2. This indicates that both Down and Proposal do not change their performance in the two simulated scenarios, that is, for these two \ac{NDN} deployments, it does not matter whether the vehicles are consuming the same data.

\begin{figure}[!htb]
	\centering
	\includegraphics[width=0.68\textwidth]{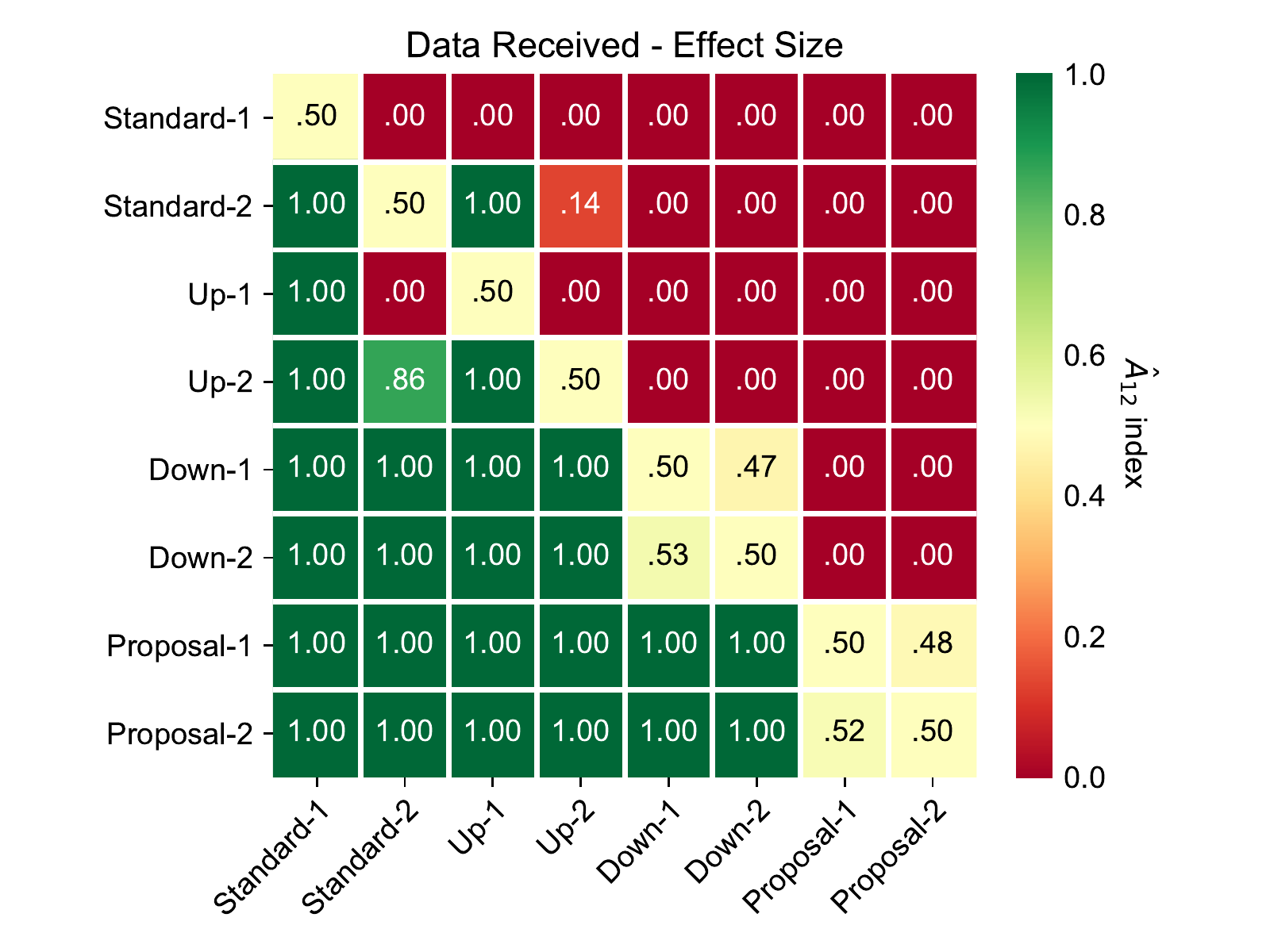}
	\caption{The Vargha and Delaney's $\hat{A}_{12}$ index.}
	\label{fig:vd}
\end{figure}

Once we concluded that there is a performance difference between the approaches evaluated, we decided to measure this difference. For that we use the $\hat{A}_{12}$ effect size test to analyze also the magnitude of the difference. This test presents an intuitive result, measuring the probability that one approach is better than another. Figure~\ref{fig:vd} shows $\hat{A}_{12}$ index for the metric of data received in vehicles in each instance evaluated of Table~\ref{tab:list_evaluated_scenarios}. The results obtained with this metric suggest that the more bytes that are transmitted via unicast, the better the performance of the NDN deployment. 

Since the Standard \ac{NDN} only transmits via broadcast it presented the worst $\hat{A}_{12}$ index, mainly in the first scenario. When we analyzed the Standard and Up deployment instances, we found that performance improves when vehicles consume the same data. This happens in 100\% of the cases when comparing Standard-2 versus Standard-1 and Up-2 versus Up-1. As shown in Table~\ref{tab:list_evaluated_scenarios}, both Up and Down deployments broadcast in only one direction. However, Down outperforms Up in 100\% of cases since transmit data packets via unicast has a greater positive impact than transmit interest packets. Our proposal outperforms the other variants in both scenarios in 100\% of cases, as it prioritizes unicast traffic in both directions. Reinforcing our premise that the more bytes transmitted via unicast, the better the performance, as it will minimize of the basic service usage of wireless networks. When observing only Down and Proposal instances, the $\hat{A}_{12}$ index for the second scenario a slight superiority despite the U-test showing no statistical difference. This slight difference is due to the requested data already in the \ac{CS} of \acp{AP} and routers.

\begin{figure}[!htb]
	\centering
	\includegraphics[width=0.68\textwidth]{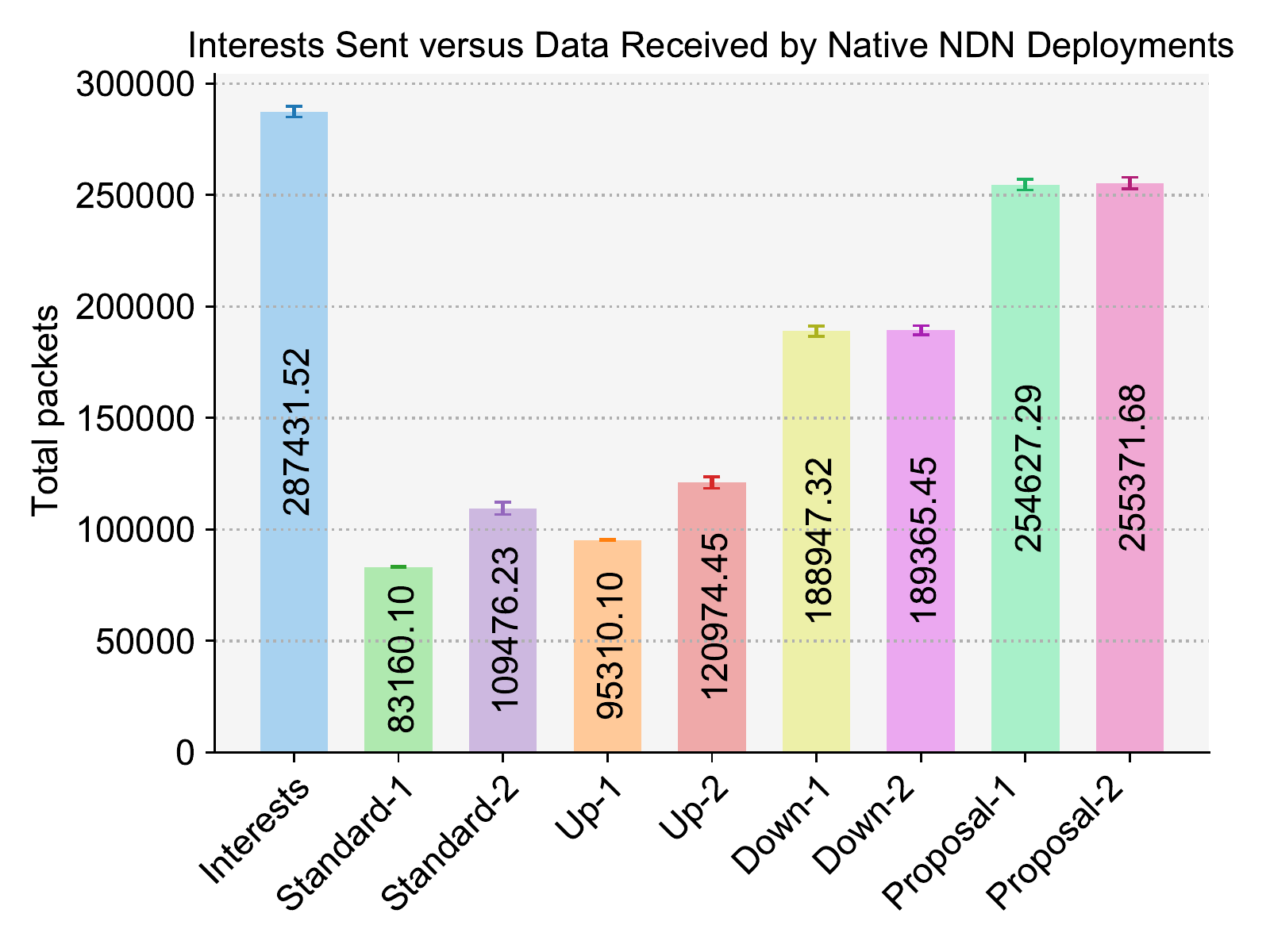}
	\caption{Relation between interest sent and data received by evaluated instances.}
	\label{fig:total_values_bar}
\end{figure}

Figure~\ref{fig:total_values_bar} shows the total number of interest packets sent and the data received in each evaluated instance. The number of interest packets sent is the same, regardless of the instance. Therefore, only one bar has been placed on the chart with this information. It is possible to observe that there is a considerable superiority of Standard-2 over Standard-1, around 31.65\% of more data received, with Standard-1 reaching only 28.93\% of requests satisfied, against 38.09\% reached by Standard-2. There is also a superiority when we compare Up-2 with Up-1, around 26.93\% of more data received, with Up-1 reaching only 33.16\% of requests satisfied, against 42.09\% reached by Up-2. Thus, it confirms the importance of vehicles requesting the same content in the Standard and Up. This superiority does not exist when looking at Down-2 versus Down-1 and Proposal-2 versus Proposal-1. Both Down instances reached around 66\% of satisfied requests. Finally, the instances of our proposal performed better than all others, reaching values close to 89\% of satisfied requests, receiving more than 200\% of data received than Standard deployments instances.

\begin{figure}[!htb]
	\centering
	\includegraphics[width=0.68\textwidth]{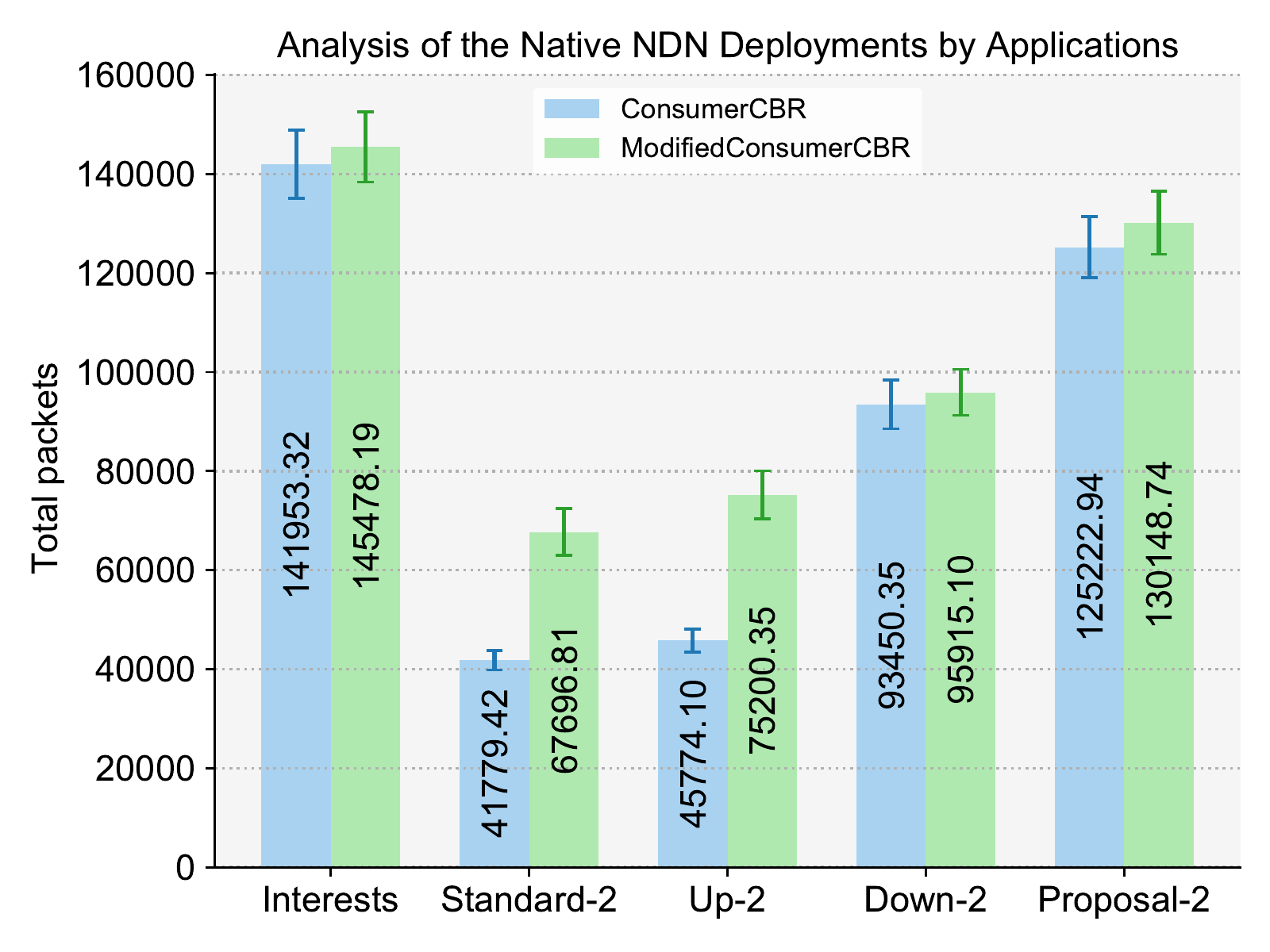}
	\caption{Relation between the interest sent and the data received by the applications (only second scenario).}
	\label{fig:total_values_by_app}
\end{figure}

This difference is explained by the fact that the wireless network standard offers a basic service of communication for broadcast transmissions~\cite{IEEE802.11-2021}, as a consequence, the performance of this traffic is reduced. Figure~\ref{fig:total_values_by_app} also shows this analysis of the relationship between interests and data, but by application, which is why it contains instances only from the second scenario. In our experiments the vehicles running \textit{ConsumerCbr} requests distinct data, while the vehicles running \textit{ModifiedConsumerCbr} requests same data. Hence, we compared NDN deployments with these two types of traffic. 

In the Standard deployment, the \textit{ConsumerCbr} application had only 29.43\% of requests satisfied and \textit{ModifiedConsumerCbr} increased to just 46.53\%. This shows that even when the same data are requested, Standard \ac{NDN} has difficulties in performing satisfactory use of the available resources on wireless networks. While in the Up deployment got a little better, even so, the \textit{ConsumerCbr} application had only 32.25\% of requests satisfied and \textit{ModifiedConsumerCbr} increased to just 51.69\%. In the Down deployment both applications reached values close to 66\%, still far from ideal. Finally, our proposal reached values close to 89\% in both applications, surpassing the other implementations. All of these results showed that the excessive use of broadcast transmissions in the link layer by Native \ac{NDN} is not scalable and reduces throughput in wireless networks. Therefore, we consider essential a layer-2 address mapping mechanism to Native \ac{NDN}, as proposed in this work.

\section{Conclusions}\label{sec:conclusions}
In this paper, we implemented a new variant of \ac{NDN} with a built-in layer-2 address mapping mechanism to minimize the excessive use of broadcast at the link layer without changing the behavior of \ac{NDN} packets at the network layer. To deploy our solution we used ndnSIM 2.8 simulator and we conducted a comparative evaluation with four Native \ac{NDN} deployments in an \ac{IoV} context considering three Wi-Fi 6 hotspots, with vehicles performing handovers along the avenue. Our vehicular traffic has been based on real traces, and from this, we propose two scenarios. In the first scenario, all vehicles request distinct data, while half of the vehicles request the same data in the second scenario.

Our proposal outperforms the Standard \ac{NDN} in the evaluated scenarios, reaching values close to 89\% of satisfied requests, receiving more than 200\% of data received than Standard deployments instances, which in the best case reached only 38.09\% of satisfied requests. The results obtained indicate that Standard \ac{NDN} has serious limitations in achieving high throughput. The main reason is due to Standard \ac{NDN} does not have a layer-2 address resolution mechanism. Unlike our proposal that has this functionality.

The problem is that contention-based wireless technologies such as Wi-Fi 6 may suffer from broadcast storms or degradation of transmission rates due to the switching to basic service~\cite{IEEE802.11-2021}. Since the Standard \ac{NDN} does not know the layer-2 addresses, it is unable to create unicast traffic. Unless the address is manually set in the \ac{NFD}~\cite{Afanasyev2018a} for each face, which is obviously not a practical alternative. Thus, the only alternative to Native \ac{NDN} is to transmit all packets via layer-2 broadcast, even if the network layer only sends to a single face. ndnSIM 2.8~\cite{Mastorakis2017a} has been developed in just that way. Therefore, we consider that our proposal can be an advance in the compatibility of Native \ac{NDN} with current wireless network technologies.

\section*{Acknowledgments}
This work was partially supported by the National Council for Scientific and Technological Development (CNPq) (Grant No. 312831/2020-0).

\bibliographystyle{sbc}
\bibliography{mybibliography.bib}

\begin{thebibliography}{}

\bibitem[{Afanasyev, A. \textit{et al}.} 2018]{Afanasyev2018a}
{Afanasyev, A. \textit{et al}.} (2018).
\newblock {NFD Developer's Guide}.
\newblock {Technical Report NDN-0021}.

\bibitem[Anastasiades et~al. 2016]{Anastasiades2016}
Anastasiades, C., Weber, J., and Braun, T. (2016).
\newblock {Dynamic Unicast: Information-centric multi-hop routing for mobile
  ad-hoc networks}.
\newblock {\em Computer Networks}, 107:208--219.
\newblock Mobile Wireless Networks.

\bibitem[Arcuri and Briand 2011]{Arcuri2011}
Arcuri, A. and Briand, L. (2011).
\newblock {A Practical Guide for Using Statistical Tests to Assess Randomized
  Algorithms in Software Engineering}.
\newblock In {\em 2011 33rd International Conference on Software Engineering
  (ICSE)}, pages 1--10.

\bibitem[{Coutinho} et~al. 2018]{Coutinho2018}
{Coutinho}, R. W.~L., {Boukerche}, A., and {Yu}, X. (2018).
\newblock {A Novel Location-Based Content Distribution Protocol for Vehicular
  Named-Data Networks}.
\newblock In {\em 2018 IEEE Symposium on Computers and Communications (ISCC)},
  pages 01007--01012.

\bibitem[CTTU 2019]{cttu2019recife}
CTTU (2019).
\newblock {Open Data of Vehicle Traffic from Recife--Brazil}.
\newblock Available in:
  http://dados.recife.pe.gov.br/dataset/velocidade-das-vias-quantitativo-por-velocidade-media-2019.

\bibitem[de~Sena and Dias 2022]{Sena2022}
de~Sena, Y. A. B.~L. and Dias, K.~L. (2022).
\newblock {Native versus Overlay-based NDN over Wi-Fi 6 for the Internet of
  Vehicles}.
\newblock In Jiang, D. and Song, H., editors, {\em {Simulation Tools and
  Techniques. SIMUtools 2021}}, volume 424, pages 51--63, Cham. Springer
  International Publishing.

\bibitem[de~Sena et~al. 2020]{Sena2020}
de~Sena, Y. A. B.~L., Dias, K.~L., and Zanchettin, C. (2020).
\newblock {DQN-AF: Deep Q-Network based Adaptive Forwarding Strategy for Named
  Data Networking}.
\newblock In {\em 2020 IEEE Latin-American Conference on Communications
  (LATINCOM)}, pages 1--6.

\bibitem[Duarte et~al. 2019]{Duarte2019}
Duarte, J.~M., Braun, T., and Villas, L.~A. (2019).
\newblock {MobiVNDN: A distributed framework to support mobility in vehicular
  named-data networking}.
\newblock {\em {Ad Hoc Networks}}, 82:77--90.

\bibitem[Gomez et~al. 2018]{rfc8352}
Gomez, C., Kovatsch, M., Tian, H., and Cao, Z. (2018).
\newblock {Energy-Efficient Features of Internet of Things Protocols}.
\newblock RFC 8352.

\bibitem[{Grassi} et~al. 2014]{Grassi2014}
{Grassi}, G., {Pesavento}, D., {Pau}, G., {Vuyyuru}, R., {Wakikawa}, R., and
  {Zhang}, L. (2014).
\newblock {VANET via Named Data Networking}.
\newblock In {\em 2014 IEEE Conference on Computer Communications Workshops
  (INFOCOM WKSHPS)}, pages 410--415.

\bibitem[{IEEE SA} 2021]{IEEE802.11-2021}
{IEEE SA} (2021).
\newblock {IEEE Standard for Information Technology -- Telecommunications and
  Information Exchange between Systems - Local and Metropolitan Area
  Networks--Specific Requirements - Part 11: Wireless LAN Medium Access Control
  (MAC) and Physical Layer (PHY) Specifications}.
\newblock {\em {IEEE Std 802.11-2020 (Revision of IEEE Std 802.11-2016)}},
  pages 1--4379.

\bibitem[Khorov et~al. 2019]{Khorov2019}
Khorov, E., Kiryanov, A., Lyakhov, A., and Bianchi, G. (2019).
\newblock {A Tutorial on IEEE 802.11ax High Efficiency WLANs}.
\newblock {\em {IEEE Communications Surveys Tutorials}}, 21(1):197--216.

\bibitem[Kietzmann et~al. 2017]{Kietzmann2017}
Kietzmann, P., G\"{u}ndo\u{g}an, C., Schmidt, T.~C., Hahm, O., and
  W\"{a}hlisch, M. (2017).
\newblock {The Need for a Name to MAC Address Mapping in NDN: Towards
  Quantifying the Resource Gain}.
\newblock In {\em 4th ACM Conference on Information-Centric Networking (ICN
  '17)}, pages 36--42.

\bibitem[Li et~al. 2015]{Li2015a}
Li, M., Pei, D., Zhang, X., Zhang, B., and Xu, K. (2015).
\newblock {NDN Live Video Broadcasting over Wireless LAN}.
\newblock In {\em 2015 24th International Conference on Computer Communication
  and Networks (ICCCN)}, pages 1--7.

\bibitem[Liang et~al. 2020]{Liang2020}
Liang, T., Pan, J., Rahman, M.~A., Shi, J., Pesavento, D., Afanasyev, A., and
  Zhang, B. (2020).
\newblock {Enabling Named Data Networking Forwarder to Work Out-of-the-Box at
  Edge Networks}.
\newblock In {\em 2020 IEEE International Conference on Communications
  Workshops (ICC Workshops)}, pages 1--6.

\bibitem[Lopez et~al. 2018]{Lopez2018}
Lopez, P.~A., Behrisch, M., Bieker-Walz, L., Erdmann, J., Flötteröd, Y.-P.,
  Hilbrich, R., Lücken, L., Rummel, J., Wagner, P., and Wiessner, E. (2018).
\newblock {Microscopic Traffic Simulation using SUMO}.
\newblock In {\em {2018 21st International Conference on Intelligent
  Transportation Systems (ITSC)}}, pages 2575--2582.

\bibitem[Mastorakis et~al. 2017]{Mastorakis2017a}
Mastorakis, S., Afanasyev, A., and Zhang, L. (2017).
\newblock {On the Evolution of ndnSIM: an Open-Source Simulator for NDN
  Experimentation}.
\newblock {\em SIGCOMM Comput. Commun. Rev.}, 47(3):19--33.

\bibitem[Nour et~al. 2019]{Nour2019}
Nour, B., Li, F., Khelifi, H., Moungla, H., and Ksentini, A. (2019).
\newblock {Coexistence of ICN and IP Networks: An NFV as a Service Approach}.
\newblock In {\em 2019 IEEE Global Communications Conference (GLOBECOM)}, pages
  1--6.

\bibitem[{ns-3} 2021]{ns3}
{ns-3} (2021).
\newblock {ns-3 Network Simulator Website}.
\newblock Available in: https://www.nsnam.org/.

\bibitem[Rothmuller and Barker 2020]{Rothmuller2020}
Rothmuller, M. and Barker, S. (2020).
\newblock {IoT - The Internet of Transformation 2020}.
\newblock pages 1--8, Basingstoke, UK.
\newblock Juniper Research.

\bibitem[Shi et~al. 2017]{Shi2017}
Shi, J., Newberry, E., and Zhang, B. (2017).
\newblock {On Broadcast-based Self-Learning in Named Data Networking}.
\newblock In {\em 2017 IFIP Networking Conference (IFIP Networking) and
  Workshops}, pages 1--9.

\bibitem[Virtanen et~al. 2020]{2020SciPy-NMeth}
Virtanen, P., Gommers, R., Oliphant, T.~E., Haberland, M., Reddy, T.,
  Cournapeau, D., Burovski, E., Peterson, P., Weckesser, W., Bright, J., {van
  der Walt}, S.~J., Brett, M., Wilson, J., Millman, K.~J., Mayorov, N., Nelson,
  A. R.~J., Jones, E., Kern, R., Larson, E., Carey, C.~J., Polat, {\.I}., Feng,
  Y., Moore, E.~W., {VanderPlas}, J., Laxalde, D., Perktold, J., Cimrman, R.,
  Henriksen, I., Quintero, E.~A., Harris, C.~R., Archibald, A.~M., Ribeiro,
  A.~H., Pedregosa, F., {van Mulbregt}, P., and {SciPy 1.0 Contributors}
  (2020).
\newblock {{SciPy} 1.0: Fundamental Algorithms for Scientific Computing in
  Python}.
\newblock {\em {Nature Methods}}, 17:261--272.

\bibitem[Wang et~al. 2020]{Wang2020}
Wang, X., Wang, Z., and Cai, S. (2020).
\newblock {Data Delivery in Vehicular Named Data Networking}.
\newblock {\em IEEE Networking Letters}, 2(3):120--123.

\bibitem[Wu et~al. 2018]{Wu2018}
Wu, F., Yang, W., Fan, Z., and Tian, K. (2018).
\newblock {Multicast Rate Adaptation in WLAN via NDN}.
\newblock In {\em 2018 27th International Conference on Computer Communication
  and Networks (ICCCN)}, pages 1--8.

\bibitem[Zhang et~al. 2014]{Zhang2014}
Zhang, L., Afanasyev, A., Burke, J., Jacobson, V., claffy, k., Crowley, P.,
  Papadopoulos, C., Wang, L., and Zhang, B. (2014).
\newblock {Named Data Networking}.
\newblock {\em SIGCOMM Comput. Commun. Rev.}, 44(3):66--73.

\end{thebibliography}

\end{document}